\documentclass[a4paper,11pt]{article}
\usepackage{pos}

\title{Comprehensive Measurement Forecasts of the EeV Neutrino-Nucleon Cross Section with Cosmic Neutrinos at IceCube-Gen2}
 \ShortTitle{UHE neutrino-nucleon cross section measurement forecast}

\author*[a]{Victor B. Valera}
\author[a]{Mauricio Bustamante}
\author[b]{Christian Glaser}

\affiliation[a]{Niels Bohr International Academy, Niels Bohr Institute, University of Copenhagen, \\ DK-2100 Copenhagen, Denmark}

\affiliation[b]{Department of Physics and Astronomy, Uppsala University,\\ Uppsala, SE-752 37, Sweden}

\emailAdd{vvalera@nbi.ku.dk}
\emailAdd{mbustamante@nbi.ku.dk}
\emailAdd{christian.glaser@physics.uu.se}

\abstract{The investigation of neutrino interactions with matter serves as a valuable tool for understanding the fundamental structure of nucleons and potentially uncovering novel physics phenomena. To date, the neutrino-nucleon cross section has been examined across a range of energies spanning from a few hundred MeV to PeV. However, the pursuit of ultra-high-energy (UHE) cosmic neutrinos, surpassing 100 PeV in energy, holds the promise of further advancements. In the next 10-20 years, UHE neutrino telescopes, currently in the planning stage, may ultimately succeed in their detection. This article presents pioneering and comprehensive estimation forecasts for the ultra-high-energy neutrino-nucleon cross section, with a specific focus on the employment of neutrino radio-detection within the IceCube-Gen2 experiment. The study incorporates cutting-edge methodologies in UHE neutrino flux prediction, neutrino propagation within the Earth, radio detection techniques, and the treatment of background data to facilitate accurate cross section measurement projections. Assuming the successful detection of at least a few tens of UHE neutrino-induced events over a 10-year period, IceCube-Gen2 could achieve, for the first time, the measurement of the cross section at center-of-mass energies of approximately $\sqrt{s} \approx 10$--100 TeV. Furthermore, if the number of events exceeds one hundred, the precision of the cross section measurement could be comparable to its corresponding theoretical prediction.}

\ConferenceLogo{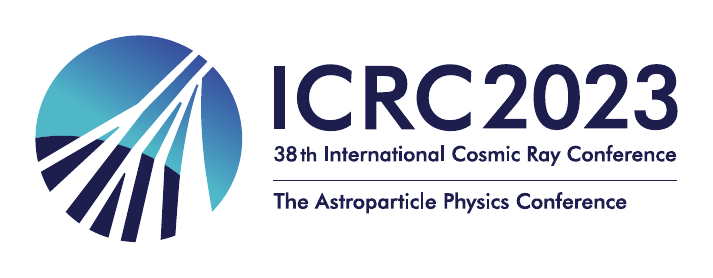}

\FullConference{%
38th International Cosmic Ray Conference (ICRC2023)\\
  26 July - 3 August, 2023\\
  Nagoya, Japan}

\begin{document}
\maketitle

\section{Introduction}

Neutrino interactions with matter offer valuable insights into particle physics by revealing the inner structure of nuclei and nucleons and potentially uncovering new physics phenomena. The probing power of neutrinos increases with higher neutrino energy. Currently, experimental knowledge of high-energy neutrino-matter interactions, specifically, the neutrino-nucleon ($\nu N$) cross section, extends up to PeV energies, the highest detected thus far for neutrinos. However, further exploration of the $\nu N$ cross section at higher energies holds great potential, which is currently limited by the capabilities of existing detectors.

\begin{figure*}[t]
 \centering
 \includegraphics[width=0.9\textwidth]{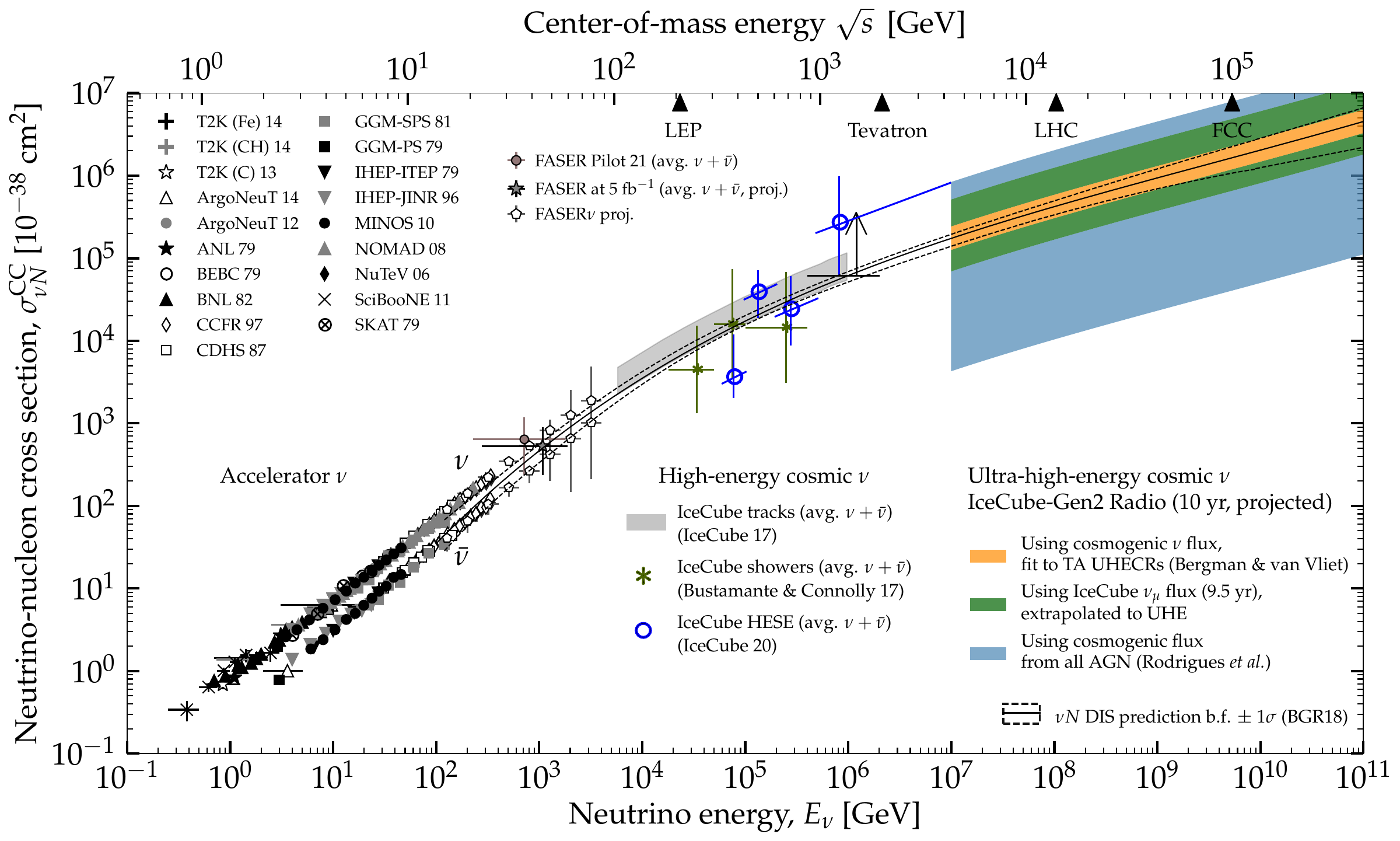}
 \caption{\label{fig:panorama}Neutrino-nucleon ($\nu N$) charged-current (CC) cross section, measurements and predictions.  Sub-TeV measurements are from accelerator-neutrino experiments. Few-TeV measurements will be covered by the upcoming FASER$\nu$ accelerator-neutrino experiment.  TeV--PeV measurements use high-energy astrophysical neutrinos detected by the IceCube neutrino telescope~\cite{IceCube:2017roe, Bustamante:2017xuy, IceCube:2020rnc}. We forecast cross-section measurements above 100~PeV after 10 years of exposure. Figure taken from Ref.~\cite{Valera:2022ylt}, adapted for Ref.~\cite{Ackermann:2022rqc}.}
\end{figure*}

Figure~\ref{fig:panorama} illustrates the current landscape of $\nu N$ cross section measurements. Precise measurements have been achieved in the energy range of approximately 100 MeV to 350 GeV using neutrino beams from particle accelerators. Although the planned FASER$\nu$ experiment will reach energies in the TeV range~\cite{FASER:2019dxq}, no neutrino beam exists or is planned beyond TeV energies. Consequently, measurements in the TeV-PeV range~\cite{IceCube:2020rnc, Bustamante:2017xuy} have relied on high-energy cosmic neutrinos detected by the IceCube neutrino telescope. While these are the most energetic neutrinos observed to date, they are not the most energetic ones predicted. Ultra-high-energy (UHE) cosmic neutrinos with energies of 100 PeV and above were predicted over fifty years~\cite{Berezinsky:1969erk} ago but have remained elusive due to their low predicted flux~\cite{Greisen:1966jv, Zatsepin:1966}. However, the development of new neutrino telescopes within the next 10--20 years holds promise for detecting UHE neutrinos even with a low flux~\cite{Ackermann:2022rqc}. The discovery of UHE neutrinos would have significant implications for astrophysics, particularly in understanding the origins of ultra-high-energy cosmic rays, as well as for particle physics, enabling tests of fundamental physics in a new energy regime and advancing cross-section measurements.

In our analysis~\cite{Valera:2022ylt}, we make the first detailed estimates of the sensitivity of the planned radio component of IceCube-Gen2~\cite{IceCube-Gen2:2020qha} to measuring the UHE neutrino-nucleon cross section. IceCube-Gen2 is among the largest UHE neutrino telescopes under consideration, and in advanced stages of planning. We use state-of-the art ingredients at all stages: in the flux predictions~\cite{IceCube:2020wum, IceCube:2021uhz, Heinze:2019jou, Fang:2013vla, Padovani:2015mba, Fang:2017zjf, Muzio:2019leu, Rodrigues:2020pli, Anker:2020lre, Muzio:2021zud}, computation of the attenuation~\cite{Garcia:2020jwr}, and estimates of the detector response~\cite{Glaser:2019cws, Glaser:2019rxw}, including energy and angular uncertainties. To account for the significant uncertainty surrounding UHE neutrino predictions, our cross-section forecasts consider various flux sizes and energy spectra from the literature. Figure~\ref{fig:panorama} demonstrates that, under optimistic flux scenarios, IceCube-Gen2 has the potential to measure the UHE $\nu N$ cross section with an accuracy of around $50\%$ compared to theoretical predictions. See Ref.~\cite{Valera:2022ylt} for a more detailed description of our methods and results. For related studies see Ref.~\cite{Esteban:2022uuw}

\section{Sensitivity to $\nu N$ cross section with cosmic neutrinos}

In our cross-section measurement forecasts, we rely on estimates of the expected number of neutrino-induced events detected by IceCube-Gen2. To understand the sensitivity to the cross section, we consider the number of detected neutrinos as a function of their energy and zenith angle. This estimation incorporates factors such as the diffuse neutrino flux, the $\nu N$ cross section, neutrino attenuation during Earth propagation, and the average number density of nucleons encountered by the neutrino inside the Earth. However, a simplified expression is used only for insight and not for producing our results:
\begin{equation}
 \label{equ:event_rate_simple}
 N_\nu(E_\nu, \theta_z) 
 \propto 
 \Phi_\nu(E_\nu) 
 \sigma(E_\nu)
 e^{-L(\theta_z)/L_{\nu N}(E_\nu, \theta_z)} \;.
\end{equation}
The full treatment of neutrino propagation and detection, which is employed in our analysis, is detailed in Sections~\ref{sec:propagation} and \ref{sec:event_rate}, respectively.

The number of events depends on the $\nu N$ cross section in two ways according to Eq.~(\ref{equ:event_rate_simple}). During propagation, the cross section affects the exponential attenuation of the neutrino flux as it passes through the Earth. Higher neutrino energies and longer travel distances lead to stronger attenuation. During detection, the cross section influences the event rate proportionally, as a larger cross section increases the chances of detecting a neutrino at the detector. The sensitivity of neutrino telescopes to the high-energy $\nu N$ cross section arises from the interplay between these two effects.

To extract the cross section, we examine the angular distribution of detected events. Downgoing events, where the neutrinos arrive from above the detector, experience negligible attenuation, resulting in mild sensitivity to the cross section due to the degeneracy between the diffuse flux and the cross section. Upgoing events, where the neutrinos arrive from well below the horizon, encounter strong attenuation, leading to a low event rate. For horizontally and nearly horizontally arriving neutrinos, the balance between attenuation and flux survival enables the detection of events. At ultra-high energies, neutrinos can only arrive from a few degrees around the horizon, and they are known as Earth-skimming neutrinos. By combining events from all directions, the degeneracy between the diffuse flux and the cross section in downgoing events is broken. 

\section{The ultra-high-energy neutrino flux at Earth}\label{sec:uhe_neutrinos}

Ultra-high-energy neutrinos, characterized by energies of 100 PeV and above, are expected to originate from interactions involving ultra-high-energy cosmic rays (UHECRs) at the EeV scale. These interactions can occur either within the cosmic accelerators that act as their sources or during their journey towards Earth. Neutrinos produced within the accelerators are referred to as \textit{astrophysical} neutrinos, while those created during the extragalactic propagation of UHECRs are known as \textit{cosmogenic} neutrinos. Due to uncertainties surrounding UHECR properties, there is a wide range of predicted neutrino flux normalizations and energy spectrum shapes. In our analysis~\cite{Valera:2022ylt}, we consider a diverse set of benchmark flux models from the literature to account for this variability.

Cosmic accelerators are expected to generate a population of non-thermal UHECRs characterized by a power-law energy spectrum. The interaction of these UHECR protons with matter ($pp$) or radiation ($p\gamma$) leads to the production of short-lived $\Delta(1232)$ resonances that subsequently decay into charged pions. These decays result in neutrino production, such as $\pi^+ \to \mu^+ + \nu_\mu$, followed by subsequent decay processes. On average, each final-state neutrino carries approximately $5\%$ of the energy of the parent proton.

For our forecasts we perform the statistical analysis on 12 different UHE neutrino flux models in the literature~\cite{IceCube:2020wum, IceCube:2021uhz, Heinze:2019jou, Fang:2013vla, Padovani:2015mba, Fang:2017zjf, Muzio:2019leu, Rodrigues:2020pli, Anker:2020lre, Muzio:2021zud}. For a detailed description of each of these models see Ref.~\cite{Valera:2022wmu}.

\section{Ultra-high-energy neutrino propagation inside the Earth}\label{sec:propagation}
To precisely estimate the neutrino-induced event rates, advanced Monte Carlo simulations are employed to propagate neutrinos within the Earth. We utilized a state-of-the-art code called \textsc{NuPropEarth}~\cite{Bertone:2018dse} for this purpose. \textsc{NuPropEarth} considers the dominant contributions from neutrino-nucleon deep inelastic scattering (DIS) interactions, including neutral-current (NC) and charged-current (CC) interactions, while also accounting for other interaction channels and regeneration effects. The code incorporates the downscattering of neutrinos to lower energies through NC interactions and the replacement of neutrinos with charged leptons through CC interactions. Additionally, it incorporates phenomena such as $\bar{\nu}_e$ scattering on atomic electrons via the Glashow resonance, $\nu_\tau$ regeneration, energy losses of intermediate tauons. The internal matter density profile of the Earth used is the Preliminary Reference Earth Model.

Having propagated the neutrino fluxes from the Earth's surface to the simulated surface of the detector, the resulting neutrino fluxes at the detector are used to calculate neutrino-induced event rates at the detector.

\section{Event-rate estimation in the radio component of IceCube-Gen2}\label{sec:event_rate}
Ultra-high-energy neutrinos are detected through the radio emission of the particle showers that they trigger. Upon reaching the detector volume, UHE neutrinos scatter off nucleons in ice, producing showers of high-energy particles. As these showers propagate, an excess of electrons accumulates at the front of the shower, emitting coherent radio pulses known as Askaryan radiation~\cite{Askaryan:1961pfb}. The advantage of using radio signals is that they experience little attenuation in ice, allowing for the construction of a sparse underground array of radio antennas over a large volume, to detect the potentially low flux of UHE neutrinos. The radio array of IceCube-Gen2 adopts this strategy. However, our methods can be adapted to other UHE neutrino telescopes.

To compute realistic event rates at the radio array of IceCube-Gen2, we employ a detailed procedure based on state-of-the-art simulations~\cite{Valera:2022ylt, Valera:2022wmu}. In brief, UHE neutrinos with energy $E_\nu$ interact with nucleons at rest in the detector volume after propagating through the Earth. The resulting particle shower has an energy $E_{\rm sh}$, which is a fraction of the energy of the parent proton. The energy of the shower depends on the interaction channel and flavor of the neutrino involved. The detector response, represented by its effective volume, depends on the shower energy and the direction of the incoming neutrino. We employ simulations using \textsc{NuRadioReco}~\cite{Glaser:2019rxw} and \textsc{NuRadioMC}~\cite{Glaser:2019cws} tools to generate the effective volume, accounting for the generation and propagation of Askaryan radiation in ice.

\begin{figure*}[t!]
 \centering
 \includegraphics[width=\textwidth]{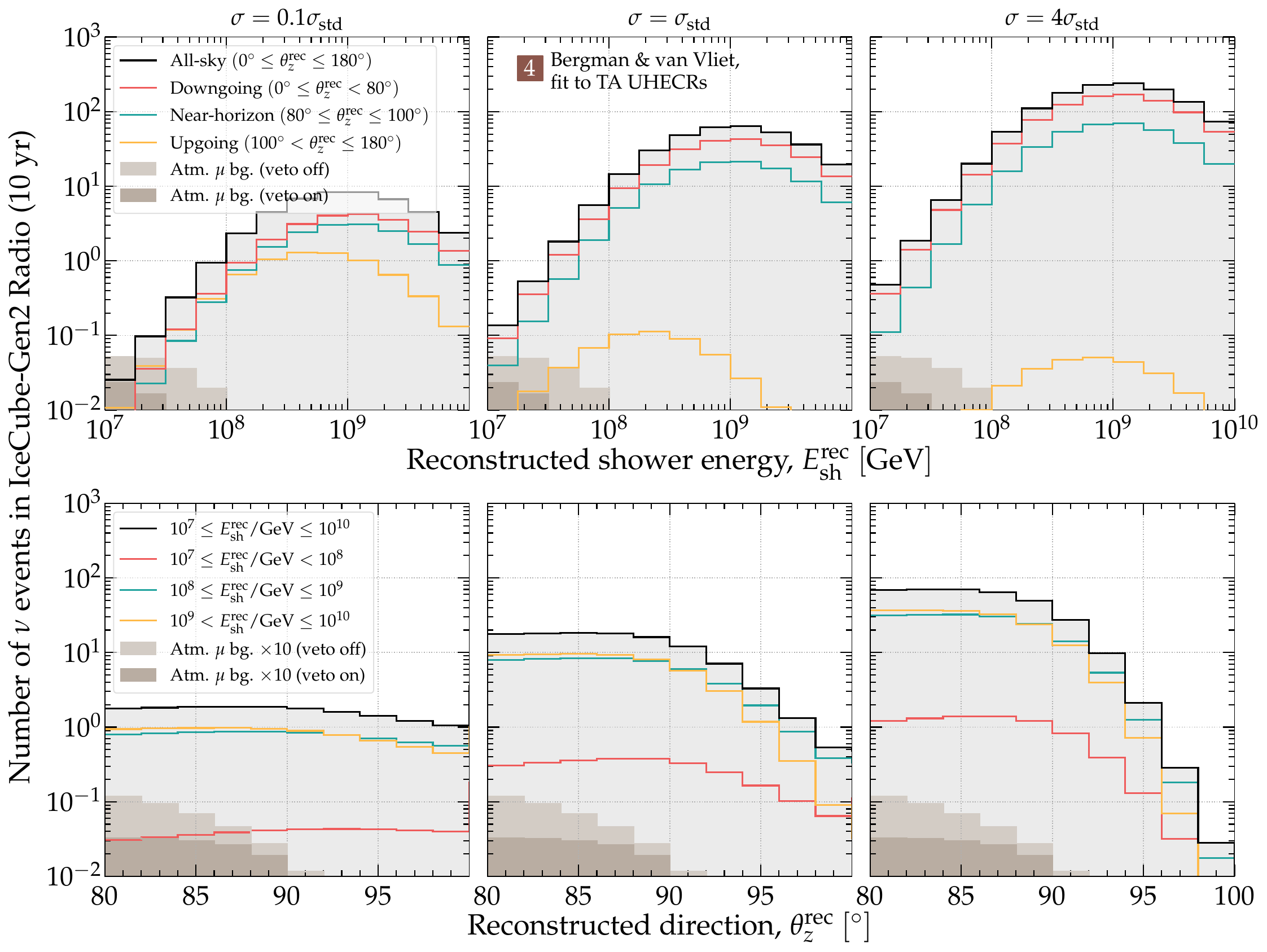}
 \caption{\label{fig:event_rate}Mean expected distribution of neutrino-initiated showers in the radio component of IceCube-Gen2 after 10 years of exposure.  For this plot, we assume the UHE diffuse neutrino flux model in Ref.~\cite{Anker:2020lre}. Each column shows a different choice of the $\nu N$ DIS cross section that affects the propagation of neutrinos through the Earth and their detection.  We include the background of showers due to atmospheric muons, which factors into our statistical analysis, computed using the hadronic interaction model {\sc Sybill 2.3c}~\cite{Fedynitch:2018cbl}.  In our analysis, we use exclusively the background after the application of the surface veto. Figure taken from Ref.~\cite{Valera:2022ylt}.}
\end{figure*}

\section{Statistical methods}

To assess the capability of IceCube-Gen2 to recover the true values of the $\nu N$ cross section and flux normalization, we employ a statistical procedure. For this purpose, we generate both \textit{true} and \textit{test} shower samples according to Sec.~\ref{sec:event_rate} and compared, considering a non-neutrino background composed by high-energy atmospheric muons as well.

The true shower sample represents the mean expected sample for IceCube-Gen2 and is constructed using the central values of the BGR18 DIS calculation for the cross section ($\sigma_{\rm std}$) and the nominal flux normalization ($\Phi_{0, {\rm std}}$) evaluated at a reference energy of $E_{\nu,0} = 10^8$ GeV. Test shower samples are generated using different values of the cross section ($\sigma_{\nu_\alpha N}$) and a different flux normalization ($\Phi_0$), which deviate from the true values.

The deviations from the true $\nu N$ DIS cross section are parametrized by the factor $f_\sigma = \sigma / \sigma_{\rm std}$. The modifications of the cross section are assumed to be energy-independent and common across neutrino flavors and interactions with protons and neutrons. The true shower sample is computed with $f_\sigma = 1$, while multiple test shower distributions are generated by allowing $f_\sigma$ to vary from 0.01 to 100. The change in $f_\sigma$ affects the neutrino flux reaching the detector through modifications in neutrino propagation through the Earth and the cross section at detection. However, it only impacts the $\nu N$ DIS cross section, while other sub-leading neutrino interaction channels during propagation remain unchanged.

Similarly, deviations from the true flux normalization are parametrized by $f_\Phi = \Phi_0 / \Phi_{0, {\rm std}}$, where $\Phi_{0, {\rm std}}$ is specific to each flux model. Varying $f_\Phi$ shifts the flux normalization while keeping the shape of the neutrino spectrum, flavor composition, and $\nu$-$\bar{\nu}$ content unchanged. The true shower sample is computed with $f_\Phi = 1$, and test shower distributions are generated by allowing $f_\Phi$ to vary over a range.

The statistical analysis involves comparing the test shower samples with the true sample, evaluating the impact of deviations in the cross section and flux normalization. We conduct these procedures to determine the capability of IceCube-Gen2 to recover the true values in a realistic experimental scenario. We employ a Bayesian statistical analysis to produce forecasts in this study.

We employ a binned Poisson likelihood in our Bayesian model. We use 12 reconstructed shower energy bins equally spaced between $10^7-10^{10}$~GeV in logarithmic space, while for the reconstructed arrival direction we use 10 bins in the most relevant region of the zenith angle, between $80^\circ$ and $100^\circ$, in addition to three bin for the regions $[0^\circ, 80^\circ ]$, $[100^\circ, 110^\circ ]$, and $[110^\circ, 180^\circ ]$.

Given the \textit{true shower} distribution, the compatibility with the mean test shower distribution for a particular choice of $\boldsymbol\theta = (f_\sigma, f_\Phi)$ is evaluated using the likelihood function defined as,
\begin{equation}
 \label{equ:likelihood_partial}
 \mathcal{L}_{ij}(\boldsymbol\theta)
 =
 \prod_{i=1}^{N_{E_{\rm sh}^{\rm rec}}} \prod_{j=1}^{N_{\theta_z^{\rm rec}}} \frac{
 \bar{N}_{{\rm test},ij}(\boldsymbol\theta)
 ^{N_{{\rm obs},ij}}
 e^{-\bar{N}_{{\rm test},ij}(\boldsymbol\theta)}
 }
 {
 N_{{\rm obs},ij}!
 } \;.
\end{equation}

We use wide priors for $\log_{10} f_\sigma$ and $\log_{10} f_\Phi$ to allow for exploration of large deviations from the true values of $f_\sigma = 1$ and $f_\Phi = 1$. The priors are flat to avoid introducing bias. The range for $\log_{10} f_\sigma$ spans from -1 to 2, covering large possible modifications of the cross section, while the range for $\log_{10} f_\Phi$ varies for each flux model and is determined based on the flux at $E_{\nu,0} = 10^8$~GeV and allows for a relative change of four orders of magnitude in each case. The posterior probability distribution is then obtained by multiplying the likelihood function with the prior distribution.

Finally, the forecast results are obtained by maximizing the posterior distribution with respect to $f_\sigma$ and $f_\Phi$ to find their best-fit values and credible intervals.

\section{Results and conclusions}

In this study we employed state-of-the-art methods and tools at every stage. The BGR18 deep-inelastic-scattering (DIS) calculation served as the baseline for the $\nu N$ cross section. We considered various flux models of UHE neutrinos, encompassing extrapolations from IceCube TeV--PeV fluxes, cosmogenic neutrinos, neutrinos from cosmic-ray sources, and self-consistent joint production models. Each neutrino species was treated separately. We simualted the neutrino propagation through Earth {\sc NuPropEarth}, while the detection method focused on radio-detection of neutrino-induced showers using {\sc NuRadioMC} and {\sc NuRadioReco} tools.To account for uncertainties in the predicted neutrino flux, we adopted a Bayesian statistical approach, providing mean sensitivity results averaged over multiple observed event rate realizations.

The analysis suggests that the UHE $\nu N$ cross section could be measured within $50\%$ of the BGR18 prediction within 10 years of IceCube-Gen2 exposure, given the detection of a sufficient number of neutrino-induced showers ($N_\nu \approx 100$). In an optimistic scenario ($N_\nu \approx 300$), comparable precision could be achieved within 5 years. The primary source of uncertainty in the forecasts is the UHE neutrino flux. Nonetheless, the level of precision attained will enable testing of the standard $\nu N$ cross section prediction, investigating non-linear effects in nucleon structure, exploring color-glass condensates and sphalerons, and probing deviations introduced by new-physics models.

Although in this proceeding we focus on our nominal result, that is the measurement forecast of the $\nu N$ cross section, in Ref.~\cite{Valera:2022ylt} we show further results and explore in detail how our results depend on the different experimental parameters. With the methods developed in this study we can quantify the discovery potential of the radio component of IceCube-Gen2 in terms of the detector characteristics. This includes the angular resolution, the energy resolution, number and type of antennas, sensitivity to near-horizontal events, and exposure time. This allows us to provide informed recommendations for the upcoming generation of UHE neutrino detectors.

Although the forecasts are comprehensive, further work is necessary. It includes estimating the impact of cosmic-ray-induced shower backgrounds, improving modeling of reconstruction uncertainties, and extending the analysis to infer both the neutrino flux normalization and energy spectrum shape from data. The calculation framework allows for easy incorporation of these improvements, and ongoing work is being conducted in these areas.


%
%
%

\end{document}